\documentclass[10pt,conference]{IEEEtran}
\IEEEoverridecommandlockouts

\usepackage{hyperref}
\usepackage[utf8]{inputenc}
\usepackage{graphicx}
\usepackage{textcomp}
\usepackage{xcolor}
\usepackage{booktabs}
\usepackage{tikz}

\usepackage{soul}
\usepackage{xspace}
\usepackage{url}
\usepackage{listings,multicol,multirow}
\usepackage[caption=false]{subfig}  
\usepackage[export]{adjustbox}
\usepackage{breakurl} 
\usepackage{algorithm}
\usepackage[noend]{algpseudocode}
\usepackage{diagbox}
\usepackage{fixfoot}
\usepackage{enumerate}
\usepackage[shortlabels]{enumitem}
\usepackage{amsmath} 
\usepackage{cleveref}

\usepackage{balance}

\makeatletter 
\newcommand{\linebreakand}{%
  \end{@IEEEauthorhalign}
  \hfill\mbox{}\par
  \mbox{}\hfill\begin{@IEEEauthorhalign}
}
\makeatother 

\newcommand{\platfromName}{\textsc{JetBrains Academy}\xspace}

\newcounter{insight}
\newcommand{\insight}[1]{\refstepcounter{insight}
	\begin{center}
		\framebox{
			\begin{minipage}{0.93\columnwidth}
				{\bf RQ\arabic{insight} Summary:} \textit{#1}
			\end{minipage}
		}
	\end{center}
}

\def\BibTeX{{\rm B\kern-.05em{\sc i\kern-.025em b}\kern-.08em
    T\kern-.1667em\lower.7ex\hbox{E}\kern-.125emX}}
\begin{document}

\title{In-IDE Programming Courses: Learning Software Development in a Real-World Setting}

\author{\IEEEauthorblockN{Anastasiia Birillo}
\IEEEauthorblockA{\textit{JetBrains Research} \\
Belgrade, Serbia \\
anastasia.birillo@jetbrains.com}
\and
\IEEEauthorblockN{Ilya Vlasov}
\IEEEauthorblockA{\textit{JetBrains Research} \\
Belgrade, Serbia \\
ilya.vlasov@jetbrains.com}
\and
\IEEEauthorblockN{Katsiaryna Dzialets}
\IEEEauthorblockA{\textit{JetBrains} \\
Munich, Germany \\
katsiaryna.dzialets@jetbrains.com}
\and
\linebreakand 
\IEEEauthorblockN{Hieke Keuning}
\IEEEauthorblockA{\textit{Utrecht University} \\
Utrecht, Netherlands \\
h.w.keuning@uu.nl}
\and
\IEEEauthorblockN{Timofey Bryksin}
\IEEEauthorblockA{\textit{JetBrains Research} \\
Limassol, Cyprus \\
timofey.bryksin@jetbrains.com}
}

\maketitle

\begin{abstract}
While learning programming languages is crucial for software engineers, mastering the necessary tools is equally important. To facilitate this, JetBrains recently released the JetBrains Academy plugin, which customizes the IDE for learners, allowing tutors to create courses entirely within IDE.

In this work, we provide the first exploratory study of this learning format. We carried out eight one-hour interviews with students and developers who completed at least one course using the plugin, inquiring about their experience with the format, the used IDE features, and the current shortcomings. Our results indicate that learning inside the IDE is overall welcomed by the learners, allowing them to study in a more realistic setting, using features such as debugging and code analysis, which are crucial for real software development. With the collected results and the analysis of the current drawbacks, we aim to contribute to teaching students more practical skills.
\end{abstract}


\section{Introduction}
\label{sec:intro}

\begin{figure*}[t]
    \centering
    \includegraphics[width=0.87\linewidth]{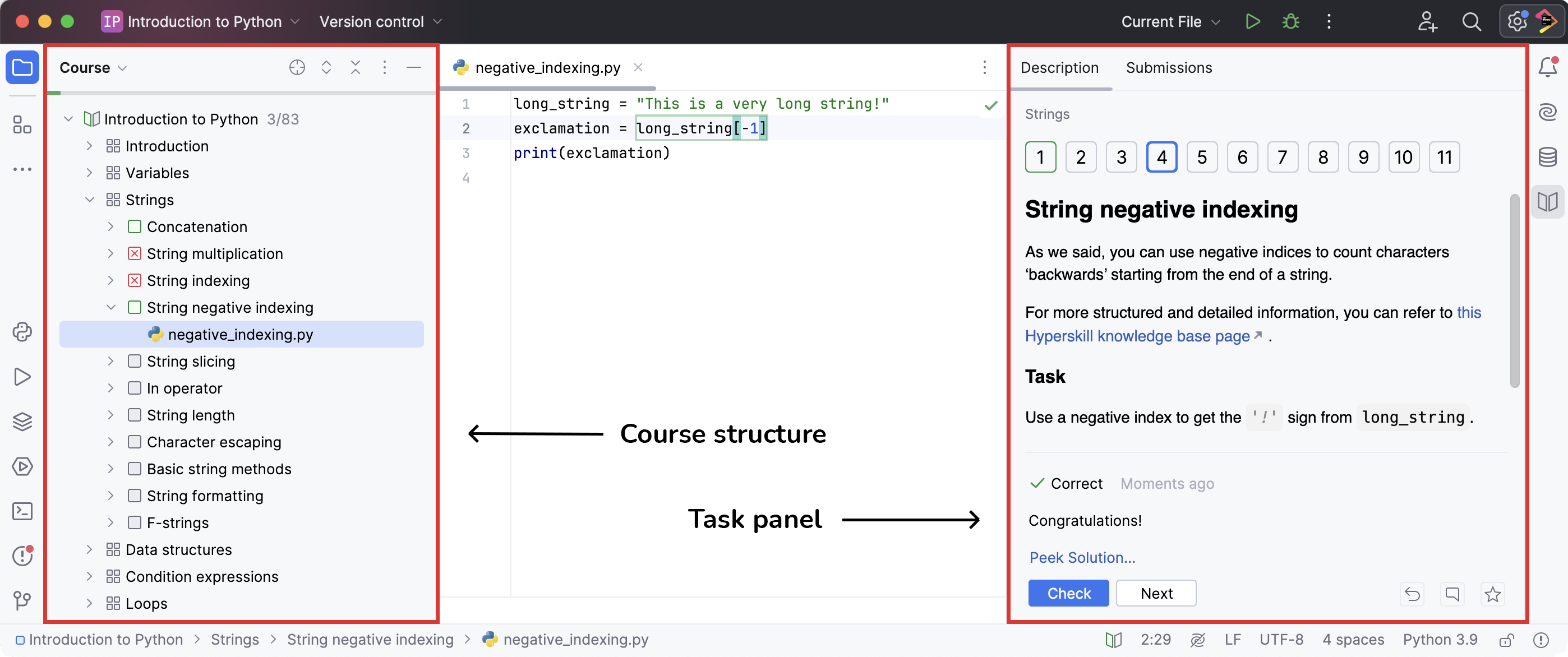}
    \vspace{-0.3cm}
    \caption{An example of the JetBrains Academy Plugin with the \textit{Introduction to Python} course. Left: course structure and course progress. Right: task panel with theory and task description.}
     \vspace{-0.3cm}
    \label{fig:examples}
\end{figure*}

Nowadays, massive open online courses (MOOCs) constitute an important way to learn programming~\cite{guerrero2023mooc}, which became especially apparent during the pandemic~\cite{cramarenco2023student, impey2021moocs}.
MOOCs on platforms like Udemy~\cite{udemy}, edX~\cite{edx}, Alison~\cite{Alison}, FutureLearn~\cite{futureLearn}, and Codecademy~\cite{codecademy} are usually designed to combine theory and practice to teach programming skills.
Being web platforms, they are typically equipped with online editors, which helps to avoid the problems of installing and setting up the necessary software. This allows students to focus on understanding new topics and writing the correct code without worrying about the rest~\cite{yulianto2017harmonik, kusumaningtyas2020online}.

Some platforms, such as Hyperskill~\cite{hyperskill}, allow users to switch to an integrated development environment (IDE) to solve more complex problems. 
With this hybrid approach, students learn theory within the platform and have the opportunity to choose which development environment to use --- a simple web editor or a professional IDE.
This enables them to learn how to use professional tools, which are an integral part of industrial programming~\cite{birillo2024bridging}. However, some students might not switch to the IDE and thus miss these crucial skills.

JetBrains~\cite{jetbrains} is an international provider of IDEs, \textit{e.g.}, IntelliJ IDEA, PyCharm, and others. At JetBrains, we believe it is beneficial for students to learn development tools together with programming concepts from the very beginning of their learning journey.
A few years ago, we released the \platfromName plugin~\cite{jetbrains-academy-plugin}, which allows creating courses \textit{fully within the IDE}, giving students the ability to both learn theory and solve practical exercises in the same coding environment~\cite{birillo2024bridging}. 
This approach ensures that students learn to use IDEs, making them even more prepared for future software development jobs.

There are many studies on how people solve coding tasks within an IDE~\cite{blogg4626326improving, mangaroska, fischer2023addressing}, but they largely focus on the IDE as just an editor and on editor-based actions. In this study, we aim to focus on the IDE as a real-world tool with all of its features, where students study theory and solve tasks in one place. 
Our study addresses three main research questions:

\begin{enumerate}[leftmargin=1.1cm,start=1,label={\bfseries RQ\arabic*:}]
    \item What are the benefits of the moving the learning process to a professional IDE?
    \item Which specific IDE features do students use when solving programming tasks inside an IDE?
    \item What are the disadvantages of the current format of the in-IDE courses, and what needs to be added to them?
\end{enumerate}

To answer these questions, we conducted eight one-hour interviews with students with different backgrounds. The results indicate that the in-IDE format is highly beneficial for students, enabling them to learn theory and complete exercises in the same environment. Furthermore, we identified the key issues and potential enhancements that could be implemented to optimize the learning experience for students.
We believe that these results will facilitate enhancements to the in-IDE study format within the JetBrains Academy plugin, in conjunction with other analogous plugins~\cite{orion-plugin, app-courses-plugin, code-tour}. 

\section{Related Work}

One of the problems that students have after completing their programming education in MOOCs is not knowing how to use IDEs and various IDE-based tools, such as debuggers~\cite{radermacher2014investigating}. For this reason, the way students behave in IDEs is of special interest to researchers, with many studies investigating IDE actions that students perform in the process of solving programming tasks~\cite{dyke2011aspects, fuchs-monitoring-android, pullar2023g, hundhausen2017ide}. 

Dyke~\cite{dyke2011aspects} tracked Eclipse IDE usage to identify novice programmers in need of support, 
however, the set of collected actions was limited to running and debugging processes, editing files, and using quick fixes. They found a direct correlation between the use of non-editorial actions and final grades, indicating the importance of teaching these actions.

Fuchs et al.~\cite{fuchs-monitoring-android} collected data of IDE and browser interactions during the development of a mobile application. This study shows that students often use file editing operations such as \textit{open}, \textit{copy}, and \textit{paste}, but rarely use IDE features, \textit{e.g.}, external libraries import list optimization or rename refactorings, which are actively used by professional developers~\cite{mastropaolo2023automated}.

In a recent paper, Pullar-Strecker et al.~\cite{pullar2023g} tested three methods for evaluating student performance models on data from introductory programming courses. This study mainly looked at file editor actions. These actions represent editor-based features and do not indicate what other, more powerful, features were used, such as refactoring or debugging.

While some of these studies investigate different aspects of learning and IDE usage, most of them focus on code editor actions. This paper provides the first exploratory study of the in-IDE learning format on the example of \platfromName plugin, providing insights about using that format for course creators, plugin developers and researchers.

\section{The \platfromName plugin}

\platfromName plugin is a plugin for JetBrains IDEs, which allows to create programming courses inside the IDE~\cite{birillo2024bridging} and share them, either privately or publicly. 
The plugin provides different types of course assignments, such as coding tasks, guided projects, quizzes, and theory tasks.
The plugin adapts the IDE window by adding a course structure panel, the course progress bar, task description panel with theory, explanations and non-coding types of assignments, and a check button for running tests predefined by the course author (see \Cref{fig:examples}).
A detailed description of the plugin functionality is provided elsewhere~\cite{birillo2024bridging}.

Currently, there are over 50 courses available for the plugin. This study is based on the following courses: \textit{Java for Beginners}~\cite{java-for-beginners}, \textit{Introduction to Python}~\cite{introduction-to-python}, \textit{C++ Basics}~\cite{cpp-basics}, \textit{Kotlin Onboarding: Introduction}~\cite{kotlin-onboarding-introduction}, and \textit{Kotlin Onboarding: Object-Oriented programming}~\cite{kotlin-onboarding-oop}. 
These courses provide an overview of the language and are designed for learners with limited or no prior experience in programming. Additionally, there are courses tailored to more experienced students, such as \textit{e.g.}, \textit{Python Libraries --- NumPy}~\cite{python-numpy}. Furthermore, the \textit{IDE Features Trainer}~\cite{ide-feature-trainer} offers insights into the IDE’s built-in features, including the use of inspections and refactorings.

\section{Methodology}

To carry out the qualitative analysis of the usefulness of the \platfromName plugin, we conducted structured interviews. 
More specifically, we first carried out a qualification survey to get a list of possible participants and then invited selected people for the main interview part.

\subsection{Qualification Survey}

The survey aimed to ensure the diversity of opinion, inviting people with different levels of experience.
It consisted of eight questions about age, programming experience, and IDE usage. The full survey is provided in the supplementary materials~\cite{appendix}.

We recruited participants by sharing the survey on LinkedIn. After a week, filtering out respondents who did not use the plugin or just started using it, we collected \textbf{121 responses}.
We grouped the respondents into three \textit{usage groups} according to how long they used the plugin: \textit{a couple of weeks} (10 out of 121), \textit{1--2 months} (35 out of 121), and \textit{3 months or more} (76 out of 121). In each group, we additionally grouped participants into \textit{occupation sub-groups}: \textit{Student}, \textit{Working student}, \textit{Freelancer}, \textit{Partially employed by a company or organization}, \textit{Fully employed by a company or organization}, or \textit{Unemployed}.  
We randomly selected 10--11 participants in each of the \textit{usage groups}, trying to get people from all \textit{occupation sub-groups}, to be invited for an interview. In total, we sent invitations to 32 people, out of which 29 booked an interview.

\subsection{Interview Process}

The interviews were conducted in English and recorded using Google Meet~\cite{google-meet}. All the participants consented to be recorded. The reward for each successful interview was a \$150 Amazon gift card or a one-year subscription to JetBrains' \textit{All Products Pack}, as the participant chose. 

\textbf{Successful interviews}. Of the 29 scheduled interviews, 12 were successful (\textit{i.e.}, were completed and resulted in the full data collection). Among the remaining, several participants did not come, and others showed themselves to be untrustworthy interviewees. At the beginning of the interview, we asked several basic questions about the plugin and finished the interview early with participants who demonstrated any lack of understanding. Of the 12 completed interviews, we consider the first four to be pilot since we used them to adjust the interview questions. Thus, we had \textbf{eight full interviews} that we used for analysis. The description of the participants from the final sample can be found in~\Cref{table:interview:participants}.

\begin{table}[t]
\centering
\caption{\label{table:interview:participants} Overview of interviewees.}
\vspace{-0.2cm}
\begin{tabular}{lccccc}
\toprule 
 & \textbf{Country} & \textbf{Occupation} &\textbf{Plugin experience}\\ \midrule  
\textbf{P1} &  Canada  & Student & A few weeks  \\ 
\textbf{P2} &  United States  & Student & 3 months+\\ 
\textbf{P3} &  United States  & Student & 1–2 months\\  
\textbf{P4} &  United States  & Freelancer & 1–2 months \\ 
\textbf{P5} &  Poland  & Unemployed & 3 months+  \\
\textbf{P6} &  United States  & Student & 1–2 months \\  
\textbf{P7} &  Luxembourg  & Fully employed & 3 months+ \\  
\textbf{P8} &  Cyprus  & Student & 3 months+  
\\ \bottomrule
\end{tabular}
\end{table}

\textbf{Questions}. Our structured interviews were divided into three parts. The first part focused on the participants themselves, \textit{e.g.}, which courses they tried or how long they had been using the IDEs. The second \textit{main} part was about their experience with the \platfromName plugin --- what problems they had and what they liked the most, as well as what IDE features they used during the learning process. The last part focused on artificial intelligence (AI) technologies and how they can be used for learning purposes~\cite{appendix}. 

\subsection{Interview analysis}

To answer the research questions, we performed \textit{transaction log analysis (TLA)}~\cite{jansen2006search}, which consists of \textit{data preparation} and \textit{data analysis} stages. 
For the \textit{data preparation} stage, we made transcripts of the interviews using the automated Otter~\cite{otter} tool, followed by the first author correcting unrecognized words. Then, we summarized the respondents’ answers. 

\textit{Data analysis} consisted of \textit{categorization} and \textit{open coding}. Categorization groups responses by our research questions, while open coding labels them accordingly.

The \textit{categorization} was conducted by the first two authors. They independently categorized each interviewee's responses according to the research questions, and then discussed the disagreements and reached consensus. The discussions were held after all participants were labeled.

The \textit{open coding} was conducted by the same two authors. For each research question, they independently proposed labels that comprise the different answers to the question. They had a discussion round to make the final list of labels to be used as a summary for each research question. For each question, only one round of discussion was needed to reach a consensus.  

\section{Results} 

\subsection{RQ1: Benefits of Using an IDE During Learning}

Overall, our study shows that even novice programmers without much experience in working with IDEs find the in-IDE format to be helpful for many reasons. 
Most of the benefits are related to working within an industry IDE and not the \platfromName plugin directly, which fully relates to the original purpose of the plugin to move the whole learning process into a professional environment.

\textbf{Consistent environment.} Six out of eight participants said that one of the main advantages was that they did not have to change environments between studying and coding. P6 said: \textit{``It has a lot of good educational content, you can actually code and learn simultaneously on the same platform.''}

\textbf{IDE features.} More specific to the IDE, six participants said they use various IDE features---\textit{e.g.}, debugging, spellchecker, or navigation---to make the programming process easier. For example, P4 mentioned auto-formatting: \textit{"Auto-formatting ensures that my code has the consistent style that I'm using, it enhances readability and maintainability."} 

\textbf{Personalization and plugins.} Five participants shared their experience of personalizing the environment for their own use, as the IDE allows you to install extra plugins or change the IDE settings to suit your needs. This is mostly used by more advanced learners who already have enough IDE expertise, industrial experience, or a mentor who can advise on how they can use the IDE in a more efficient way. In particular, participants mentioned the \textit{Rainbow Brackets}~\cite{rainbow-brackets} plugin that allows to highlight various types of brackets and indents in different colors. P1 said: \textit{``He [dad] actually told me that I should start with this one [Rainbow Brackets].  $<$...$>$ This plugin actually helps me to read the code by showing different levels of parentheses in different colors.''}
Also, two more experienced participants used the \textit{Key Promoter X}~\cite{keypromoter} plugin, which suggests an IDE shortcut for the last used action. P5 mentioned: \textit{``Key Promoter is absolutely amazing. It's very fast to use only the keyboard and not be forced to wave your hand between the keyboard and the mouse.''}

\textbf{AI plugins.} Five participants mentioned AI plugins as an additional source of information, \textit{e.g.}, providing information about functions or a set of arguments, or even helping to find an error in the program. For example, P6 said: \textit{``Sometimes I copy all the code into it [ChatGPT plugin] and ask it to tell me what's wrong with this code.''}

\textbf{Ease of use.} Three participants highlighted that they found this format of studying to be easy to use in the context of interaction with the plugin and the IDE.  One participant had difficulties the first time, but when doing the second course, the format was easy to use even after a break between the courses. They said: \textit{``The second [course] was way easier, because by then I had actually had my experience.''}

\textbf{Other.} Finally, participants highlighted that it was useful for them that the \platfromName plugin indicates their mistakes in theory and coding tasks (two participants). P3 put it like this: \textit{``It can detect your faults, like where you have mistakes or when you get stuck.''} Other similar comments mentioned the tracking of progress in the left tool window (two participants) and the immediate feedback (one participant).

\insight{The in-IDE format is helpful for learners because it allows them to be in the same environment throughout the entire educational process. Also, the environment can be easily customized via plugins to suit each learner's needs, which helps them to learn IDE features and familiarize themselves with the industry technologies they will need for their future jobs.}

\subsection{RQ2: Usage of Specific IDE Features}

Given that participants say that the access to IDE features is a benefit of the approach, we asked them which particular IDE features they used when completing the courses. 

\textbf{Debugger.} The most popular used feature among participants is the debugger. Six out of eight participants use it to debug their code when solving tasks. Three more advanced participants even use breakpoints with conditions. P5 commented: \textit{``It's very helpful that the breakpoint can be activated only with, for example, selected values of some argument. Because I want to avoid unnecessary breaking.''} Although the majority of participants use the debugger, no one uses all of its features, \textit{e.g.}, a window for calculating the values or expressions. Also, participants said that they often prefer to debug their code the ``old-fashioned way'', by putting print statements around their program. However, this still indicates the usefulness of the format, as it incentivizes learners to use the features that they will need in real software development.

\textbf{Code analysis.} The second most popular category of features is code analysis. Three participants use inspections to automatically detect code issues, such as \textit{incorrect use of a construction} or \textit{incorrect formatting}. Two participants say they use the automatic reformatting of the entire file according to the language style conventions. P5 said: \textit{``I use Ctrl+Alt+L combination to make my code prettier."} 
Two other participants shared their experiences of using documentation in code to understand the purpose of some classes or functions.

\textbf{Navigation.} Participants who have taken the extra course on using the IDE or have had access to more experienced colleagues, use features such as code navigation, refactoring, and shortcuts. P4 elaborated: \textit{``I use code navigation. I go to the definitions, and it makes it easier for me to explore and understand my code base''.} These features help to write better code in a more efficient way.

\insight{The main used features are the debugger and code analysis. Our interviews showed that more experienced learners use more features and these features can be reused in their work environment as well. However, despite the availability of a large number of features, it is still difficult for learners to know about them, as the current version of the plugin does not provide any guidance or training on such features.}

\subsection{RQ3: Disadvantages of the Learning Format}

\textbf{Unused features.} The first problem we discovered is that just moving the process into the IDE is not enough to make learners actually use advanced features. Even if they have enough experience with IDEs and programming, and have completed the \textit{IDE Features Trainer} course, they may use only popular features like the debugger. Even then, they do not use advanced features like conditional breakpoints. 
Currently, when students start a course with the plugin, it gives them an opportunity to learn how to use the IDE, but unfortunately this is not enough without practice during course solving. P8 admitted: \textit{``I just skip the hints, but in any case, I learned something useful from them.''} 

\textbf{Searching for help.} The next major issue is the lack of help when problems with code or the environment arise. All eight participants turned to a more experienced mentor (friend, teacher, relative) for help or used online resources such as YouTube or ChatGPT to find an answer. In general, this can even be useful, since students learn to use external resources to solve problems, however, learners often cannot ask for help immediately and spend hours searching for solutions to problems. P3 mentioned:  \textit{``There was a time I was trying to write a little code, and I got stuck. And it took me weeks to get unstuck. So I was just keeping on trying, trying, doing a lot of things.''} Participants also identified a number of key problems when working with AI tools such as ChatGPT. The most common problem mentioned by six participants was the lack of trust related to the incorrect answers or the fear of personal data breaches. P8 mentioned: \textit{``But if it [ChatGPT] deceived me once, I won't ask him a second time.''} Participants also mentioned problems such as these tools not working properly for complex programs, not working well for non-English languages, and the overall process being slower, as students have to spend more time double-checking the results.

\textbf{Setting up.} Even though the plugin itself is mostly easy to use, as mentioned in RQ1, three participants had difficulty setting  it up. 
The plugin does not allow running courses in a preinstalled environment, and participants have to fix these kind of problems manually. 
P7 explained:  \textit{``The most common thing that I can recall is not finding the Python interpreter.''}

\insight{The current version of the plugin does not provide enough help on how to use the IDE correctly. This includes the extent of using IDE features, lack of the personalized help, and the correct configuration of the course environment. All these problems should be fixed in order to provide a more useful platform for students.}
\section{Implications}

The presented analysis of the advantages and disadvantages of the in-IDE learning format can be used in the future by  \textit{course creators}, \textit{plugin developers} in various IDEs, as well as \textit{researchers} involved in computing education research.

\textbf{Course creators.} Our results showed that students overall enjoy the in-IDE format. However, with the current version, course creators should provide more information within the courses to help students improve their experience.

First, the course authors could suggest the usage of supplementary plugins to enhance the learning experience of the students: (1) \textit{Rainbow Brackets}~\cite{rainbow-brackets}, is designed to assist novice students in navigating the use of brackets and indents. This plugin can be used in courses on Java, Kotlin, Python, Rust, JavaScript, and others~\cite{rainbow-brackets-githib}. (2) \textit{Key Promoter X}~\cite{keypromoter}, which analyses the student's action and recommends an IDE shortcut to perform this action. This plugin may be particularly beneficial in courses whose objective is to instruct students in the utilization of IDE features. For example, it could be employed in courses such as \textit{Introduction to IDE Code Refactoring} in Kotlin~\cite{refactorings-kotlin} and Java~\cite{refactorings-java}.

Secondly, it may be beneficial to provide extra clarifications within the task descriptions, thus preventing students from becoming overly reliant on guidance. The course authors may consider predefining clarification hints like those in the \textit{Kotlin Onboarding: Introduction}~\cite{kotlin-onboarding-introduction}, and \textit{Kotlin Onboarding: Object-Oriented programming}~\cite{kotlin-onboarding-oop} courses.

Finally, it would be beneficial to develop additional courses that instruct students in the utilization of IDE features, such as the debugger, or code analysis with navigation. Nevertheless, incorporating assessments into these courses may prove challenging, as the course creators lack direct access to the IDE features in the tests. This issue could be addressed by utilizing the  IntelliJ Platform SDK~\cite{ij-sdk}, similar to the \textit{Introduction to IDE Code Refactoring in Kotlin}~\cite{refactorings-kotlin} course. This course employs the \textit{Kotlin Test Framework}~\cite{kotlin-test-framework} developed by JetBrains, which leverages the capabilities of a concrete syntax tree~\cite{psi} from the IntelliJ Platform SDK to ascertain whether a refactoring has been executed. The sole constraint is that it is only compatible with JVM languages.

\textbf{Plugin developers.} Together with the positive aspects, this study highlighted the current limitations of the format and the lack of guidance for students during the learning process. This need to be accounted for by future developers and added to the in-IDE learning format to the \platfromName plugin~\cite{jetbrains-academy-marketplace} and other plugins, such as \textit{A++ Courses}~\cite{app-courses-plugin} or \textit{Orion -- Artemis Programming Exercise Integration}~\cite{orion-plugin} for JetBrains IDEs and Code Tour~\cite{code-tour} for Visual Studio. 

One of the most significant challenges identified by students is the lack of individualized assistance during the learning process. However, the mere utilization of tools such as ChatGPT may not be sufficient to enhance the learning experience~\cite{aruleba2023integrating, qureshi2023chatgpt}. One potential method for plugin developers to address this issue is to develop a personalized system and integrate it into their plugin. There is a substantial body of research examining the potential of large language models (LLMs) for this purpose~\cite{liu2024teaching, liffiton2023codehelp, aruleba2023integrating, kiesler2023exploring}. In case of the JetBrains Academy plugin, we have endeavoured to employ some of these techniques and integrate them with the IntelliJ Platform SDK, with the objective of providing students with next step hints~\cite{birillo2024one}. 

Another aspect that needs improvement is the development of an IDE feature recommendation system that can support students throughout their learning journey, obviating the need for additional courses or plugins. 

\textbf{Researchers.} Our work can facilitate researchers to conduct studies in this domain, when the entire learning process is moved into professional IDEs. The research might be conducted in different directions.

One possibility is to conduct a comprehensive investigation into the specific challenges associated with in-IDE learning. This could be increasing the number of participants or diversifying the geographical locations where they reside and limit the scope of the study with a few problems. It may also prove beneficial to assess whether different strategies should be employed for course creation and plugin development, tailored to the preferences of different learner groups.

Another area for research is the question of how students can be assisted to actually \textit{learn} the IDE features, rather than \textit{only using} them with the IDE recommendations or employing them solely during a dedicated course on this purpose.

\section{Threats to Validity}

\textbf{Sample.} With eight participants, our sample for the interviews is not very large. However, this is normal for exploratory qualitative studies in the field~\cite{danielsson2023identity}. In addition, we made effort to ensure that our participants have diverse backgrounds, checked their understanding, and used pilot interviews.

\textbf{Analysis.} It is possible that our analysis of the interviews missed something. To combat this, both categorization and open-coding were carried out by the first two authors independently, with iterative discussions and reaching consensus.

\section{Conclusion}
In this paper, we presented and studied different aspects of moving the learning process to a professional IDE with the \platfromName plugin from the learners' point of view. 
Overall, our exploratory study of the in-IDE learning format clearly demonstrates its potential to improve the practical skills of learners. At the same time, the mentioned drawbacks and the desired features indicate the need for further research and development.
Fundamentally, simply moving the entire learning process into the IDE is not enough in itself --- it is crucial to integrate it well with the IDE and showcase the IDE's capabilities to the student. Luckily, the IDE's powerful internal tooling, such as static analysis, provides many opportunities for this.  We believe our findings will help to improve the in-IDE study format in general and also facilitate new research in this direction.

\bibliographystyle{IEEEtran}
\balance
\bibliography{acmart}

\end{document}